\documentclass[%
 reprint,
 amsmath,amssymb,
 aps,
pra,
]{revtex4-2}

\usepackage{graphicx}
\usepackage{dcolumn}
\usepackage{bm}
\usepackage{xcolor}
\usepackage{braket}
\usepackage{float}
\usepackage{lipsum}
\usepackage{caption}
\usepackage{soul}
\usepackage{subcaption}
\usepackage{float} 

\begin{document}
\preprint{APS/123-QED}

\title{Anomalous transparency of photons in non-Markovian coupled waveguides}

\author{J. R. Silva}
\email{jefferson.rocha@fis.ufal.br}
\author{P. A. Brand\~ao}
\email{paulo.brandao@fis.ufal.br}
 \affiliation{Instituto de F\'isica, Universidade Federal de Alagoas, Macei\'o, Brasil}





\date{\today}

\begin{abstract}
The transmission of photons through a pair of coupled single-mode waveguides is studied in detail. It is assumed that one of the waveguides is coupled to a non-Markovian reservoir, described by a Lorentzian spectrum distribution. We identified four distinct regimes for the transmission coefficient and found that, in some conditions, it is more efficient to launch the photons in the lossy waveguide to achieve high transmission. We also report on a loss-induced transparency effect that is induced by the distribution of the frequencies present in the reservoir. 
\end{abstract}

\maketitle


\section{Introduction}

Integrated photonics has experienced significant advancements in recent years, driven by the emergence of quantum technologies \cite{flamini2018photonic,wang2020integrated,bao2023very,yang2024programmable}. By using arrays of coupled waveguides, researchers have demonstrated various key quantum photonic operations and effects, including the controlled-NOT logic gate \cite{politi2008silica}, multiphoton quantum interference \cite{marshall2009laser}, on-chip multiphoton entanglement \cite{matthews2009manipulation}, quantum walks \cite{peruzzo2010quantum}, polarization-encoded qubits \cite{crespi2011integrated}, boson sampling \cite{crespi2013integrated}, Mach-Zehnder interferometry \cite{wang2014gallium}, single-photon $W$-state generation \cite{grafe2014chip}, experimental studies on quantum complexity \cite{carolan2014experimental}, Bloch oscillations \cite{lebugle2015experimental} and perfect quantum state transfer \cite{chapman2016experimental} to cite a few.

The experimental systems mentioned, and indeed every physical system in the real world, are subject to dissipation. The coupling of a quantum system to its environment can lead to decoherence, a topic of extensive and highly active research \cite{breuer2016colloquium,de2017dynamics,breuer2002theory}.  

Relevant to the present analysis is the physical system of two coupled waveguides, where one is subjected to a dissipative environment. This system has been studied experimentally by first using classical optical fields \cite{guo2009observation}. Counterintuitively, instead of reducing transmission as one might expect, increasing the loss level in one of the waveguides led to an increase in the transmission coefficient. The authors in \cite{guo2009observation} attributed this phenomenon to the Parity-Time (PT) invariant properties of the structure having exceptional points \cite{miri2019exceptional,heiss2012physics,li2023exceptional,bender1998real,longhi2018parity,bender2019pt}. A review of PT-symmetric quantum mechanics, which explains many of the observed effects in waveguide systems, has been published recently \cite{bender2024pt}.

The importance of PT symmetry in the description of quantum mechanics was suggested by C. M. Bender and S. Boettcher many years ago in the context of nonrelativistic quantum mechanics \cite{bender1998real}. They provided numerical evidence on the reality of eigenvalues of non-hermitian Hamiltonians having an exact PT symmetry \cite{bender1999pt} and later found the correct inner product necessary to maintain hermiticity, which involves a new symmetry, named $C$ by the authors \cite{bender2002complex}. Mostafazadeh was the first to point out that PT-symmetric Hamiltonians are a subclass of a more general class of pseudo-hermitian Hamiltonians \cite{mostafazadeh2002pseudo,mostafazadeh2010pseudo,mostafazadeh2003exact}.

The first experimental results revealing PT-symmetric effects made use of classical optical fields, which started by Guo and collaborators in 2009 \cite{guo2009observation} (see also \cite{ruter2010observation}). A genuine PT-symmetric phenomenon related to a quantum measurement was observed and published only six years ago \cite{klauck2019observation}. The authors considered the coincidence properties of photons arriving at the detectors after propagating in the coupled waveguide system under Markovian conditions \cite{grafe2013correlations} and measured the non-Hermitian HOM dip effect. A few theoretical proposals have been published exploring the quantum dynamics in this waveguide system, including anti-parity-time symmetry \cite{qin2021quantum} and unequal losses of the waveguides \cite{zhou2022analytical,zhou2022characterization}.

We focus here on the transmission properties of such coupled waveguides in the presence of non-Markovian environments. Thus, the average number of photons leaving the system is the main quantity of interest. The anomalous (or, loss-induced) transparency effect, verified in \cite{guo2009observation}, has been theoretically explored in the quantum realm very recently \cite{beder2024quantum}. It was found that the effect still persists at the single-photon level under Markovian conditions. Also, the transmission depends on the quantum correlation properties of the incident photons. Since it is now a common practice to engineer environments having non-Markovian properties \cite{li2020non,liu2011experimental,liu2018experimental}, one may ask how the propagation of photons is affected by its interaction with a reservoir having memory effects.

In section II(A), we describe the theoretical model by defining the Hamiltonian of the system and the Heisenberg equations of motion. Section II(B) is devoted to a general discussion of the properties of the Lorentzian environment. Section III presents the results and discussions for two initial conditions of the system that generates zero and nonzero interference terms in the transmission coefficient. In section IV we present our conclusions.

\section{Theoretical Model}

This section is divided into two subsections. Subsection A deals with the description of the Hamiltonian and the Heisenberg equations of motion. Section B is devoted to a discussion of the non-Markovian environment.

\subsection{The Hamiltonian and Heisenberg equations}

Consider a pair of single-mode coupled waveguides, $W_A$ and $W_B$, where $W_B$ is connected to a reservoir (see Fig. 1). The total system is described by the Hamiltonian $H = H_0 + H_I$, where $(\hbar = 1)$
\begin{equation}
    H_0 = \omega ( a^{\dagger}a + b^{\dagger}b) + \sum_{n}\beta_n r_n^{\dagger}r_n,
\end{equation}
and
\begin{equation}
    H_I = \kappa(a^{\dagger}b + ab^{\dagger}) + \sum_{n}\alpha(\beta_n)(b^{\dagger}r_n + br_n^{\dagger}),
\end{equation}
where $a$ ($b$) is the annihilation operator for waveguide $W_A$ ($W_B$), $\omega$ is the propagation constant, assumed equal for both waveguides, $r_n$ is the annihilation operator for the (bosonic) reservoir mode having propagation constant $\beta_n$, $\kappa$ is the coupling between waveguides $W_A$ and $W_B$ and $\alpha(\beta_n)$ is a real-valued coupling constant between waveguide $W_B$ and mode $\beta_n$ of the reservoir.

\begin{figure}[H]
    \centering
    \includegraphics[width=0.9\linewidth]{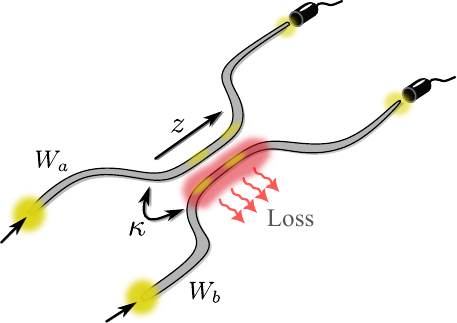}
    \caption{A pair of coupled single-mode waveguides, $W_A$ and $W_B$ having coupling constant $\kappa$ is considered. The waveguide $W_B$ is connected to a non-Markovian reservoir. The waveguides have a  constant cross-section and lie along the $z$ axis.}
    \label{fig:sys}
\end{figure}

In the Heisenberg picture, the operators $a(z)$, $b(z)$ and $r_n(z)$ evolve according to the set of coupled differential equations,
\begin{equation}\label{eq3}
    \frac{da}{dz} = -i\omega a - i\kappa b,
\end{equation}
\begin{equation}\label{eq4}
    \frac{db}{dz} = -i\omega b - i\kappa a - i\sum_n \alpha(\beta_n)r_n,
\end{equation}
\begin{equation}\label{eq5}
    \frac{dr_n}{dz} = -i\beta_n r_n - i\alpha(\beta_n)b.
\end{equation}
After substituting the integral equation of Eq. \eqref{eq5},
\begin{equation}\label{eq6}
\begin{split}
    r_n(z) = r_n(0)&e^{-i\beta_n z}\\
    &- i\alpha(\beta_n)\int_0^z e^{-i\beta_n (z-z')} b(z')\, dz',
\end{split}
\end{equation}
into Eq. \eqref{eq4} and following the usual approach of taking the continuous limit of all sumations over the reservoir modes,  $\sum_n \rightarrow \int_0^{\infty} \, d\beta \rho(\beta)$, $\rho(\beta)$ being the reservoir mode density, Eq. \eqref{eq4} turns into
\begin{equation}\label{eq7}
    \begin{split}
        \frac{db}{dz} = &-i\omega b - i\kappa a \\
        &-\int_0^{\infty}\,d\beta \, \rho(\beta)\alpha^2(\beta)\int_0^z e^{-i\beta(z-z')}b(z')\, dz'\\
        &- i \int_0^{\infty} \rho(\beta)\alpha(\beta) e^{-i\beta z}r(\beta,0)\, d\beta.
    \end{split}
\end{equation}
The positive quantity $S(\beta) = \rho(\beta)\alpha^2(\beta)$ is the spectral density of the reservoir and Eq. \eqref{eq7} can be written in a more interesting form in terms of the autocorrelation function $R(\tau)$, defined by
\begin{equation}\label{R}
    R(\tau) = \int_0^{\infty} S(\beta)e^{-i\beta \tau }\, d\beta,
\end{equation}
which is a manifestation of the Wiener-Khichin theorem \cite{goodman2015statistical}. The usual Markovian approximation assumes that $R(\tau) \propto \delta(\tau)$ so that the dynamics of $db(z)/dz$ depends only on $z$ and not $z'< z$. We keep the analysis general and do not assume the Markovian approximation in what follows. Finally, one obtains
\begin{equation}\label{eq9}
    \begin{split}
        \frac{db}{dz} = &-i\omega b - i\kappa a -R\ast b\\
        &- i \int_0^{\infty} \rho(\omega)\alpha(\beta) e^{-i\beta z}r(\beta,0)\, d\beta,
    \end{split}
\end{equation}
where $R\ast b$ is the convolution between $R$ and $b$. 

Our objective is to find $a(z)$ and $b(z)$ from the pair of coupled equations \eqref{eq3} and \eqref{eq9}. Notice that, once $b(z)$ has been found, Eq. \eqref{eq6} can be used to obtain $r(\beta,z)$, if necessary. We proceed by using the Laplace transform. Equations \eqref{eq3} and \eqref{eq9}, in Laplace domain, are given by
\begin{equation}
    sA(s) = a(0) -i\omega A(s) - i\kappa B(s),
\end{equation}
\begin{equation}
\begin{split}
    sB(s) &= b(0) -i\omega B(s) - i\kappa A(s) - M(s)B(s) \\    
    &-i\int_0^{\infty}\frac{\rho(\beta)\alpha(\beta)}{s+i\beta}r(\beta,0)\,d\beta ,
\end{split}
\end{equation}
where $A(s)$ and $B(s)$ are the Laplace transforms of the operators $a(z)$ and $b(z)$, $\text{Re}(s) > 0$ and $M(s)$ is the Laplace transform of $R(\tau)$:
\begin{equation}\label{eq12}
    M(s) = \int_0^{\infty}e^{-s\tau}R(\tau)\,d\tau.
\end{equation}
The solutions to the coupled algebraic operator equations involving $A(s)$ and $B(s)$, can be written as

\begin{equation}\label{eq13}
\begin{split}
    A(s) &= F_1(s)a(0)+G(s)b(0)\\
    &+\int_0^{\infty}H_1(\beta,s)r(\beta,0)\,d\beta
\end{split}
\end{equation}
\begin{equation}\label{eq14}
\begin{split}
    B(s) &= F_2(s)b(0)+G(s)a(0)\\
    &+\int_0^{\infty}H_2(\beta,s)r(\beta,0)\,d\beta
\end{split}
\end{equation}

To simplify notation, the parameters $F_j(s)$, $G(s)$ and $H_j(\beta,s)$ were introduced. They are defined by
\begin{equation}\label{Fjs}
    F_j(s)=\frac{s+i\omega+\delta_{1j}M(s)}{(s+i\omega)\Big[s+i\omega+M(s)\Big]+\kappa^2}
\end{equation}
\begin{equation}\label{Gs}
    G(s)=\frac{-ik}{(s+i\omega)\Big[s+i\omega+M(s)\Big]+\kappa^2}
\end{equation}
\begin{equation}
    H_1(\beta,s) = -\frac{iG(s)\rho(\beta)\alpha(\beta)}{s+i\beta}
\end{equation}
and 
\begin{equation}
    H_2(\beta,s) = -\frac{iF_2(s)\rho(\beta)\alpha(\beta)}{s+i\beta}.
\end{equation}

The inverse Laplace transform can now be applied directly to Eqs. \eqref{eq13} and \eqref{eq14} to obtain

\begin{equation}\label{eq19}
\begin{split}
    a(z) = f_1(z)a(0) &+ g(z)b(0) \\
    &+ \int_0^{\infty}h_1(\beta, z)r(\beta,0)\, d\beta, 
\end{split}
\end{equation}

\begin{equation}\label{eq20}
\begin{split}
    b(z) = f_2(z)b(0) &+ g(z)a(0) \\
    &+ \int_0^{\infty}h_2(\beta,z)r(\beta,0)\, d\beta, 
\end{split}
\end{equation}
where $f_j$, $g$ and $h_j$ are given by the respective inverse Laplace transforms of $F_j$, $G$ and $H_j$

The functions $f_j(z)$, $g(z)$ and $h_j(z)$ complete characterize the dynamics of the system in Markovian and non-Markovian situations. The integrals in Eqs. \eqref{eq19} and \eqref{eq20} are the random noise source terms, necessary to guarantee the correct commutation relations between the operators $a$ and $b$ at every $z$. 

\subsection{Lorentzian environment}

We assume that waveguide $W_B$ is coupled to an environment described by a Lorentzian distribution,
\begin{equation}
    S(\beta) = \frac{\alpha^2}{\pi}\frac{\gamma/2}{(\beta-\beta_c)^2+(\gamma/2)^2}, 
\end{equation}
where $\beta_c$ is the frequency where the Lorentzian peak is located, $\gamma$ is the FWHM of the distribution and $\alpha$ is the coupling parameter between waveguide $W_B$ and the reservoir. 

Assuming that $\gamma\ll\beta_c$, the integral in Eq. \eqref{R} can be extended to $(-\infty,\infty)$ and the autocorrelation function $R(z)$ is given by $R(z) = \alpha^2e^{-i\beta_cz}e^{-\gamma z/2}$. In the following, we consider $\beta_c = \omega$. Finally, the Laplace transform of $R(z)$ [Eq. \eqref{eq12}] is given by
\begin{equation}\label{mslorentz}
    M(s) = \frac{\alpha^2}{s+i\omega+\gamma/2}.
\end{equation}
The function $M(s)$ given by Eq. \eqref{mslorentz} must then be substituted back into Eqs. \eqref{Fjs} and \eqref{Gs} and the inverse Laplace transform taken to obtain $f_j(z)$ and $g(z)$, which completely determine the field dynamics.

\begin{figure}[H]
    \centering
    \includegraphics{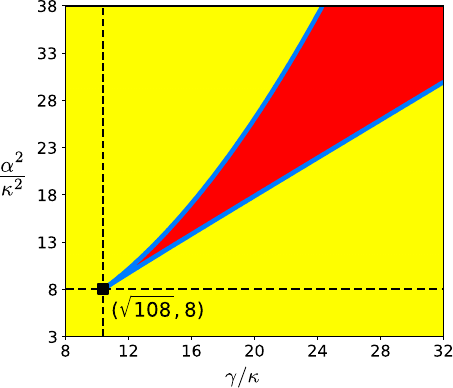}
    \caption{This figure indicates the general behavior of the roots of the polynomial \eqref{Qspol} as function of $\alpha^2/\kappa^2$ and $\gamma/\kappa$. The red region represents three distinct real roots and the yellow region indicates solutions with one real root and a pair of complex conjugate roots. The blue line represents the case of double degenerate roots and the single square point $(\sqrt{108},8)$ represents the situation where the three roots are equal.}
    \label{fig2} 
\end{figure}

It is possible to drawn some general conclusions about the behavior of $f_j(z)$ and $g(z)$ by inspecting Eqs. \eqref{Fjs} and \eqref{Gs}. Since the denominator is a cubic polynomial with real coefficients $(s \rightarrow s-i\omega)$, 
\begin{equation}\label{Qspol}
    Q(s)=s^3+\frac{\gamma}{2}s^2+(\alpha^2+\kappa^2)s+\kappa^2\frac{\gamma}{2},
\end{equation}
there are four situations to consider depending on the nature of the discriminant of $Q(s)$: (a) Three distinct real roots $c_1 \neq c_2 \neq c_3 \neq c_1$. In this case, the Laplace inversion behaves as $\xi_1e^{c_1z}+\xi_2e^{c_2z}+\xi_3e^{c_3z}.$ (b) Three identical real roots $c_1 = c_2 = c_3$. In this case, the function behaves as $(\xi_1+\xi_2z+\xi_3z^2)e^{c_1z}$. (c) Two equal roots $c_2 = c_3$ with $c_1 \neq c_2,c_3$. This time the solution is given by $\xi_1e^{c_1z}+(\xi_2+\xi_3z)e^{c_2z}$. (d) One real root $c_1$ and the other two forming a complex-conjugated pair $c_2 = c_3^*$. Only in this last situation we find that the resulting function has oscillations of the form $\xi_1e^{c_1z} +e^{\text{Re}(c_2)z} [ \xi_2\cos (z\text{Im}c_2 )+\xi_3\sin (z\text{Im}c_2 ) ] $.

Figure \ref{fig2} displays the behavior of all the roots of Eq. \eqref{Qspol} as a function of $\alpha^2/\kappa^2$ (vertical axis) and $\gamma/\kappa$ (horizontal axis). The red color region indicates the case (a) mentioned above, with three distinct real roots. The square point located at $\alpha^2/\kappa^2 = \sqrt{108}$ and $\gamma/\kappa = 8$ represents the case (b) of three identical roots. All points (except the one at the square) belonging to the blue line are represented by the case (c) of two equal real roots, and the yellow region represents the oscillatory situation labeled (d) previously.

\begin{figure}[H]   
    \includegraphics{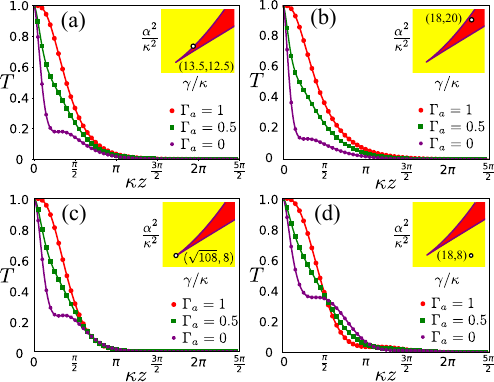}
    \caption{Transmission $T$ as a function of $\kappa z$ for (a)
    $(\gamma/\kappa,\alpha^2/\kappa^2)=(13.5,12.5)$,
    (b) 
    $(\gamma/\kappa,\alpha^2/\kappa^2)=(18,20)$, (c) 
    $(\gamma/\kappa,\alpha^2/\kappa^2)=(\sqrt{108},8)$  and (d)
    $(\gamma/\kappa,\alpha^2/\kappa^2)=(18,8)$. All plots show the relative initial population $\Gamma_a = 0$ (purple diamonds), $\Gamma = 0.5$ (green squares) and $\Gamma_a = 1$ (red circles). The insets highlight the positions in the diagram of Fig. \ref{fig2}. }
    \label{fig3}
\end{figure}

In closing this section, we would like to highlight that other equivalent approaches could be used to simulate the present model \cite{garraway1997nonperturbative,anto2021capturing,menczel2024non}. For example, in the pseudo-mode approach, our system can be mapped into another optical setup composed of three single-mode waveguides, $W_a$, $W_b$ and $W_c$, in which waveguide $W_c$ is coupled to a Markovian bath (through an effective coupling parameter $\gamma$). The Heisenberg equation for the $W_a$ mode still has the form of Eq. \eqref{eq3} while modes $W_b$ and $W_c$ satisfy
\begin{align}
    \frac{db}{dz}&=-i\omega b-i\kappa a -i\alpha c,\label{24}\\
    \frac{dc}{dz}&=-i\omega c-i\alpha b -\gamma c\label{25}.
\end{align}
The integral equation for the annihilation operator $c$ of the pseudo-mode is given by
\begin{equation}
    c(z)=c(0)e^{-(\gamma+i\omega) z}-i\alpha\int_0^zb(z')e^{-(\gamma+i\omega) (z-z')}dz',
\end{equation}
and Eq. \eqref{eq9} is recovered after substitution into the integral equation for $b$. This view could be extremely useful for the experimental implementations of structured reservoir in photonic platforms.


\section{Results and discussion}

Let us assume that the initial state of the reservoir is the vacuum state $\ket{0_R}$, so that the initial state of the total system (waveguides + reservoir) is given by
\begin{equation}
    \ket{\Psi_0} = \ket{\phi_{ab}}\otimes\ket{0_R},
\end{equation}
where $\ket{\phi_{ab}}$ indicates the initial state of the photons in both waveguides, $W_A$ and $W_B$, and $\ket{0_R} = \ket{0_{r_1}}\cdots\ket{0_{r_n}}$ represents the empty state of the oscillators in the reservoir. As mentioned in the introduction, the main quantity of interest here is the average number of photons leaving the structure. Therefore, we define the transmission $T$ as
\begin{equation}\label{defT}
    T(z)  = \frac{n_a(z) + n_b(z) }{n_a(0) + n_b(0)},
\end{equation}
where $n_a (z) = \bra{\Psi_0} a^{\dagger}(z)a(z) \ket{\Psi_0}$ and $n_b (z) = \bra{\Psi_0} b^{\dagger}(z)b(z) \ket{\Psi_0}$ are the mean photon numbers in waveguides $W_A$ and $W_B$ at distance $z$, and $a(z)$ [$b(z)$] is given by Eq. \eqref{eq19} [Eq. \eqref{eq20}]. 

After substituting Eqs. \eqref{eq19} and \eqref{eq20} into \eqref{defT}, the transmission $T$ can be written in the general form
\begin{equation}\label{T}
\begin{split}
    T&=\frac{n_a(0)}{n_a(0)+n_b(0)} (|f_1|^2+|g|^2) \\
    &+ \frac{n_b(0)}{n_a(0)+n_b(0)}(|f_2|^2+|g|^2) \\
    &+\frac{2i(f_1-f_2)g}{n_a(0)+n_b(0)}\text{Im}\left[\langle a^{\dagger}(0)b(0)\rangle\right],
\end{split}
\end{equation}
where $\langle a^{\dagger}(0)b(0)\rangle = \bra{\phi_{ab}} a^{\dagger}(0)b(0) \ket{\phi_{ab}}$. Notice the presence of an interference term (the last term) which depends on the initial photon state correlation $\langle a^{\dagger}(0)b(0)\rangle$. The transmission is independent of the noisy operators, even in the non-Markovian regime, if the reservoir is initially empty \cite{grafe2013correlations}. 

Our analysis begins by exploring Eq. \eqref{T} in cases where the interference term (the last term in the equation) vanishes. This is the case, for example, for $p$ photons initially incident at the lossless waveguide $W_A$ and $q$ photons at the lossy waveguide $W_B$, $\ket{\phi_{ab}} = \ket{p_a}\ket{q_b}$. The transmission $T$ depends only on the relative initial populations, $\Gamma_j = n_j(0)/[n_a(0) + n_b(0)]$ ($j=a,b$): 
\begin{equation}
    T = (|f_1|^2 + |g|^2)\Gamma_a + (|f_2|^2 + |g|^2)(1-\Gamma_a),
\end{equation}
where $\Gamma_a + \Gamma_b = 1$. Figure \ref{fig3} shows the transmission $T$ as a function of $\kappa z$ for three values of $\Gamma_a$, starting from the configuration where all photons are initially located inside the lossless waveguide $W_A$ ($\Gamma_a = 1)$ until they are initially located at the lossy waveguide $W_B$ ($\Gamma_a = 0$). The nonoscillatory solutions are shown in parts (a), (b) and (c) while the oscillatory solution (yellow region of Fig. \ref{fig2}) is shown in part (d). 

The plots in Fig. \ref{fig3} illustrate the  asymmetric \color{black} behavior of the system. Symmetric initial conditions ($\Gamma_a = 0$ and $\Gamma_a = 1)$ do not lead to a symmetric  dynamics. In fact, Fig. \ref{fig3}(d) clearly shows that, to achieve a high transmission coefficient $T$ in a waveguide system with total length $\kappa z$ between approximately $\pi/2 $ and $\pi$, it is more effective to inject photons into the lossy $W_B$ waveguide. 

\begin{figure}[H]   
    \includegraphics{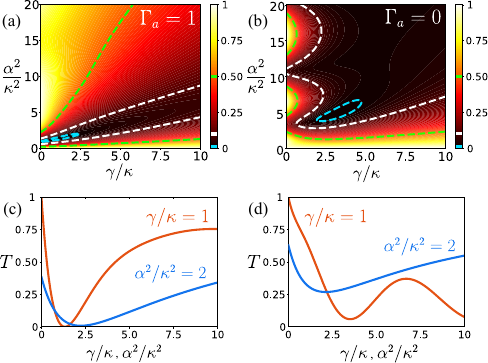}
    \caption{(a) Plot of the transmission $T$ in the $(\gamma/\kappa,\alpha^2/\kappa^2)$-plane for $\Gamma_a = 1$. (b) Same as in part (b) with $\Gamma_a = 0$. (c) 1D cross-section taken from part (a) at $\gamma/\kappa = 1$ (orange) and at $\alpha^2/\kappa^2 = 2$ (blue), displaying the increasing in transmission. (d) Same as in part (c) but for $\Gamma_a = 0$. The dashed lines in parts (a) and (b) mark constant values for $T = 0.01$ (cyan), $T = 0.1$ (white) and $T = 0.5$ (green). The propagation length of the struture is fixed at $\kappa z = 2.3$. }
    \label{fig4}
\end{figure}

To gain deep insight into the role of the coupling coefficient $\alpha$ and the non-Markovianity of the reservoir, represented by the parameter $\gamma$, on the transmission properties of the structure, Fig. \ref{fig4}(a) and (b) plots $T$ in the $(\gamma/\kappa,\alpha^2/\kappa^2)$-plane for a propagation length of $\kappa z = 2.3$. Part (a) is for $\Gamma_a = 1$ and part (b) assumes that $\Gamma_a = 0$. We thus see that the transmission $T$ strongly depends upon $\alpha$, $\gamma$ and $\Gamma_a$. In particular, note that there are oscillations of the transmission as $\gamma\rightarrow 0$ in the $\Gamma_a = 0$ case. This occurs because the reservoir has strong memory effects and interacts with the incident photons at waveguide $W_b$, impacting on their future dynamics. 

There is a very intuitive explanation for the behavior shown in Figs. \ref{fig4} (a) and (b) based on the quantum Zeno dynamics \cite{facchi2008quantum}. The interaction strength $\alpha$ between waveguide $W_b$ and the reservoir can be regarded as a constant observation amplitude between the system being observed, in this case the $W_b$ waveguide, and the observer (reservoir), which makes continuous measurements on $W_b$. In the case where all incident photons are in the lossless waveguide, $\Gamma_a = 1$, the increase in $\alpha$ induces a freezing on the initial state and most photons stay in $W_a$ during the propagation since the state of $W_b$ must remain empty \cite{beder2024quantum}. This explains the increasing values of $T$ in Fig. \ref{fig4}(a) for large $\alpha$ (ideally, $T \rightarrow 1$ in the limit  $\alpha\rightarrow\infty$). 

On the other hand, in the case where all photons are initially located at $W_b$, $\Gamma_a = 0$, the dynamics occur in two distinct regimes. For low values of $\gamma/\kappa$, where the system is highly non-Markovian (it has memory effects), there is an oscillation in transmission as $\alpha$ varies. However, increasing $\gamma/\kappa$, still under the limit of large $\alpha$, we observe that the transmission drops to zero. This implies that all photons tunnel to the reservoir with zero probability of return.

In Fig. \ref{fig4}(c), two 1D cross-section lines are plotted for the transmission shown in part (a) as a function of $\gamma/\kappa$ for fixed $\alpha^2/\kappa^2 = 2$ (blue line) and as a function of $\alpha^2/\kappa^2$ for fixed $\gamma/\kappa = 1$ (orange line). This last case is the loss-induced transparency effect reported previously (in Markovian enviroments) in Refs. \cite{guo2009observation,beder2024quantum}, where the transmission $T$ increases as the coupling constant $\alpha$ between waveguide $W_b$ and the reservoir increases. The blue line also displays an increase in transmission but for fixed $\alpha$. This implies that there is also an anomalous transparency as the frequency distribution in the reservoir changes. The same behavior is shown for $\Gamma_a = 0$ in Fig. \ref{fig4}(d). We emphasize that the two curves, in orange and blue colors, display distinct effects related to different physical properties: the interaction strength $\alpha$ (orange) and the reservoir width $\gamma$ (blue) that, to the best of our knowledge, has not been addressed previously.

\begin{figure}[H]   
    \includegraphics{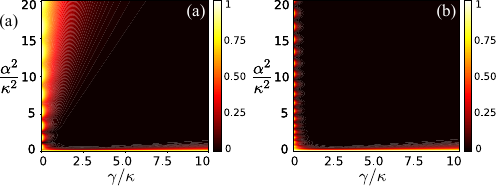}
    \caption{Transmission $T$ in the $(\gamma/\kappa,\alpha^2/\kappa^2)$-plane for (a) $\Gamma_a = 1$ and (b) $\Gamma_a = 0$. The propagation distance is fixed at $\kappa z = 15$. }
    \label{fig5}
\end{figure}

By increasing the propagation distance to $\kappa z = 15$, we obtain the plots shown in Fig. \ref{fig5}. As $\kappa z$ increases, the region of zero transmission (black color) is more evident for the case $\Gamma_a = 0$, as can be seen in Fig. \ref{fig5}(b). The nonzero values of $T$ displaying an oscillatory behavior lie at the non-Markovian limit, $\gamma \rightarrow 0$ and are expected due to the memory effects of the reservoir. 

\begin{figure}[H]
\begin{center}
    \includegraphics{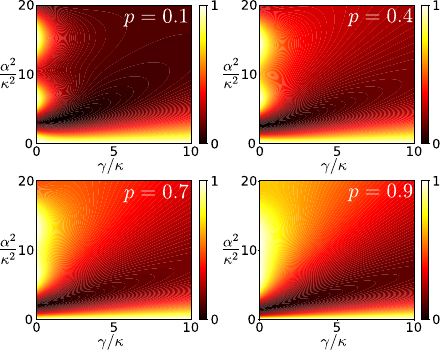}
    \end{center}
    \caption{Transmission in the ($\gamma/\kappa,\alpha^2\kappa^2$)-plane for $p = 0.1$, $0.4$, $0.7$ and $0.9$ for the initial state $\ket{\phi_{ab}} = (\sqrt{p}\ket{1_a,0_b} + e^{i\theta}\sqrt{1-p}\ket{0_a,1_b})$ with $\theta = \pi/2$. The propagation distance is fixed at $\kappa z = 2.3$ so that a comparison with Fig. \ref{fig4} can be made.}
    \label{fig6}
\end{figure}

A continuous transition between the profiles shown in parts (a) and (b) of Fig. \ref{fig4} [or parts (a) and (b) of Fig. \ref{fig5}] can be made in parametric space (for fixed $\kappa z)$ by changing the initial photonic state in the waveguides such that the last term in Eq. \eqref{T} is nonzero. So far, the number of photons incident at each waveguide is known exactly since they are given by number states. To explore situations where the last term is nonzero, let us consider the initial state $\ket{\phi_{ab}} = (\sqrt{p}\ket{1_a,0_b} + e^{i\theta}\sqrt{1-p}\ket{0_a,1_b})$, where $p\in [0,1]$ and $\theta \in [-\pi,\pi]$. This state indicates that one photon is initially sent to the waveguides but it has a probability $p$ of being found at the lossless $W_a$ and a probability $1-p$ of being found at the lossy waveguide $W_b$. It is easy to demonstrate that $\bra{\phi_{ab}}a^{\dagger}(0)b(0)\ket{\phi_{ab}} = i2\sqrt{p(1-p)}g(z)[f_1(z)-f_2(z)]\sin\theta$ and the interference term vanishes (for arbitrary $\kappa z$) in the cases where $p = 0$, $p = 1$ and $\theta = 0,\pm \pi$.

Figure \ref{fig6} displays the transition for the case where $\theta = \pi/2$ and several values of $p$ ranging from 0.1 to 0.9. One can see from these plots that there is a smooth transition between the cases considered previously that is induced by the last term in Eq. \eqref{T}. For a general initial state of the form
\begin{equation}    |\phi_{ab}\rangle=\sum_{n_a,m_b = 0}^{\infty}C_{n_am_b}|n_a,m_b\rangle,
\end{equation}
where $\ket{n_j}$ ($j = a,b$) is the number state of waveguide $W_j$ and $C_{n_am_b}$ are complex coefficients, we obtain
\begin{equation}
    \langle a^{\dagger}(0)b(0)\rangle=\sum_{n_a,m_b = 1}^{\infty}\sqrt{n_am_b}C_{n_a-1,m_b}^*C_{n_a,m_b-1},
\end{equation}
so that a sufficient condition for the interference term to vanish is given by $C_{n_a-1,m_b}^*C_{n_a,m_b-1} = 0$. This equation suggests that for interference to occur, there must be a correlation between the number of photons at each waveguide in the sense that they differ by one photon. For example, the term $\langle a^{\dagger}(0)b(0)\rangle$ is nonzero for the states $\ket{\phi_{ab}} \propto \ket{1,2} + \ket{2,1}$, $\ket{\phi_{ab}} \propto \ket{2,3} + \ket{3,2}$, $\ket{\phi_{ab}} \propto \ket{3,4} + \ket{4,3}$, etc, while it vanishes for the states $\ket{\phi_{ab}} \propto \ket{0,2} + \ket{2,0}$, $\ket{\phi_{ab}} \propto \ket{4,2} + \ket{2,4}$, $\ket{\phi_{ab}} \propto \ket{1,3} + \ket{3,1}$, etc, that differ by more than one photon.

\section{Conclusions}

We conclude by stating that there is a rich dynamics involved in the propagation of photons through non-Markovian coupled waveguides. By considering a Lorentzian model for the reservoir, we demonstrated that, in some circumstances, more light can be transmitted through the structure if the initial photons are located at the lossy waveguide. An increase in transmission was also found that depends only on the distribution of the frequencies in the dissipative environment, for fixed interaction strength between the lossy waveguide and the reservoir. A condition on the state of the incident photons to generate a nonzero contribution of the interference term in the overall dynamics was also discussed.

\begin{acknowledgments}
The authors acknowledge the financial support of CNPq (Conselho Nacional de Desenvolvimento Cient\'ifico e Tecnol\'ogico) and CAPES (Coordenação de Aperfeiçoamento de Pessoal de Nível Superior). 
\end{acknowledgments}

\appendix


\bibliography{apssamp}

\end{document}